\documentstyle[ltwol,epsfig]{article}

\def\be{\begin{equation}}
\def\ee{\end{equation}}
\def\bea{\begin{eqnarray}}
\def\eea{\end{eqnarray}}

\def\dzdzb{$D^{0}-\overline{D^{0}}$}
\def\dstarp{$D^{\star +}$}
\def\dstarwidmeas{$96\pm 4\ ({\rm stat})\pm 22\ ({\rm syst})\ {\rm
      k}e{\rm V}$}

\begin{document}

\title{Recent Results From CLEO on \dzdzb\ Mixing, $CP$ Violation in
  $D^{0}$ Decays, and \dstarp\ Width}

\author{A. B. Smith}
\address{Department of Physics, University of Minnesota, 116 Church
  St. S.E., Minneapolis, MN, 55455 USA\\E-mail: smith@hep.umn.edu}  
\author{Representing the CLEO Collaboration}


\twocolumn[\maketitle\abstracts{ 
We present preliminary results of several analyses 
searching for the effects of $CP$ violation and mixing 
in the decay of $D^0$ mesons.  
We find no evidence of $CP$ asymmetry in five different 
two-body decay modes of the $D^0$ to pairs of light pseudo-scalar mesons:
$A_{CP}(K^+ K^-) = (+0.05 \pm 2.18 \pm 0.84)\% $,
$A_{CP}(\pi^+ \pi^-) = (+2.0 \pm 3.2 \pm 0.8)\% $,
$A_{CP}(K^0_{\rm S} \pi^0) = (+0.1 \pm 1.3)\%$, 
$A_{CP}(\pi^0 \pi^0) = (+0.1 \pm 4.8)\%$, and 
$A_{CP}(K^0_{\rm S} K^0_{\rm S}) = (-23 \pm 19)\%$.
We present the first measurement of the rate of wrong-sign 
$D^0 \rightarrow K^+ \pi^- \pi^0$ decay:
$R = 0.0043^{+0.0011}_{-0.0010} \pm 0.0007$.
We also describe a measurement of the mixing parameter $y_{CP}=
{\Delta\Gamma\over 2 \Gamma}$ 
by searching for a lifetime difference between the $CP$ 
neutral $K^{-}\pi^{+}$ final state and the 
$CP$ even $K^+K^-$ and $\pi^+\pi^-$ final states.  
Under the assumption that $CP$ is conserved we find 
$y_{CP} = -0.011 \pm 0.025 \pm 0.014$.  Finally, we present our
measurement of the \dstarp\ width: \dstarwidmeas .
}]

\section{Introduction and Motivation}\label{sec:intro}
The study of mixing
in the $K^0$ and $B_d^0$ sectors has provided a wealth of information
to guide the form and content of the Standard Model.  In the framework of the 
Standard Model, mixing in the charm meson sector is predicted to 
be small~\cite{harry}, making this an excellent place to
search for non-Standard Model effects.  Similarly, measurable
$CP$ violation (CPV) phenomena in strange and 
beauty mesons are the impetus for many current and future 
experiments.
The Standard Model predictions for CPV in charm meson decay 
are of the order of $0.1\%$~\cite{Bucella}, with one recent conjecture
of nearly  
$1\%$~\cite{Bigi}.  Observation of CPV in charm meson decay exceeding
the percent level would be strong evidence for non-Standard Model processes.

A $D^0$ can evolve into a $\overline{D^0}$ through ``ordinary'' on-shell 
intermediate states, or through off-shell intermediate states, such as
those that might be present due to new physics.  We denote the
amplitude through the former (latter) states by $-iy$ $(x)$, in
units of $\Gamma_{D^0} / 2$~\cite{Mixing}.  The Standard Model contributions to
$x$ are suppressed to $|x| \approx \tan^2 \theta_C \approx 5\%$ and the
Glashow-Iliopolous-Maiani~\cite{GIM} cancellation could further
suppress $|x|$ down to $10^{-6} - 10^{-2}$.  Many non-Standard Model processes
could lead to $|x| > 1\%$.  Contributions to $x$ at this level could result
from the presence of new particles with masses as high as 100 --
1000 T$e$V~\cite{Leurer}.  Signatures of new physics include $|x| \gg |y|$ and
$CP$ violating interference between $x$ and $y$ or between $x$ and a 
direct decay amplitude.

``Wrong sign'' (WS) processes, such as $D^0 \rightarrow K^+ \pi^- \pi^0$, can proceed
directly through doubly-Cabibbo-suppressed decay (DCSD) or through
mixing and subsequent Cabibbo-favored decay (CFD).  Both DCSD and
mixing followed by CFD can contribute to the time-integrated WS
rate, $R = (r + \bar{r})/2$ and the inclusive $CP$ asymmetry 
$A = (r - \bar{r})/(r + \bar{r})$, where $r = \Gamma \left( D^0 \rightarrow 
f \right) /\Gamma \left( \overline{D^0} \rightarrow 
f \right)$, $\bar{r}$ is the charge conjugated quantity, and $f$ is a 
WS final state, such as $K^+ \pi^- \pi^0$.

The different contributions to $R$ and $A$ can be separated by 
studying the proper decay time dependence of WS final states, 
as we have done in $D^0 \rightarrow K^+ \pi^-$~\cite{kpi}, and has been done by
FOCUS~\cite{FOCUS}.  The differential WS rate
relative to the time-integrated right sign (RS) process, in time units of the
mean $D^0$ lifetime, $\tau_{D^0} = (415 \pm 4) {\rm fs}$~\cite{PDG}, is 
given by $r(t) \equiv [R_{\rm D} + \sqrt{R_{\rm D}} y^\prime t + \frac{1}{4} ({x^\prime}^2 + 
{y^\prime}^2 ) t^2 ] e^{-t}$~\cite{Treiman}.
$R_{\rm D}$ is the relative rate of DCSD, $y^\prime \equiv y \cos
\delta - x \sin \delta$, $x^\prime \equiv x  \cos \delta + y \sin
\delta$, and $\delta$ is the strong phase between the DCSD and CFD
amplitudes.  There are theoretical arguments that $\delta$ should be 
small~\cite{Wolf}, 
although this should not be taken for granted.  

A measurement of $\Gamma(D^{\star +})$ opens an important window
to non-perturbative strong physics involving heavy quarks.
The basic framework of the theory is well understood, however, there is
still much speculation -
predictions for the width range from 15~k$e$V to 
150~k$e$V~\cite{pred}.
We know the $D^{\star +}$ width is dominated by strong decays since 
the measured electromagnetic transition rate is small,
$(1.68\pm0.45)$\%~\cite{mats}.
A measurement of the width of the $D^{\star +}$
gives unique information about the
strong coupling constant in heavy-light systems.

\section{General Experimental Method}\label{genmethod}

All of the analyses discussed herein, unless otherwise stated,
use the same data set and reconstruction techniques described
below.  The data set was accumulated between February 1996 and
February 1999 and corresponds to $9.0\  {\rm fb}^{-1}$ of $e^+ e^-$ collision
data at $\sqrt{s} \approx 10.6$ G$e$V provided by the Cornell Electron
Storage Ring (CESR).  The data were recorded by the CLEO II
detector~\cite{cleoii} upgraded with the installation of a silicon vertex
detector (SVX)~\cite{svx} and by changing the drift chamber gas from
an argon-ethane mixture to a helium-propane mixture.  The 
upgraded configuration is referred to as CLEO II.V.      
        
The Monte Carlo simulation of the CLEO II.V detector is based on
GEANT~\cite{GEANT}, and simulated events are processed in the same manner
as the data.

The $D^0$ candidates are reconstructed through the decay sequence
$D^{\star +} \rightarrow D^0 \pi^+_{\rm s}$~\cite{charge}.
The charge of the slow pion ($\pi^+_{\rm s}$) tags the
flavor of the $D^0$ candidate at production.  The charged daughters
of the $D^0$ are required to leave hits in the SVX and these tracks
are constrained to come from a common vertex in three dimensions.
The trajectory of the $D^0$ is projected back to its
intersection with the CESR luminous region to obtain the $D^0$
production point.  The $\pi^+_{\rm s}$ is refit with the requirement that
it come from the $D^0$ production point, and the confidence level of the $\chi^2$ 
of this refit is used to reject background.  

The energy release in the $D^{\star +} \rightarrow D^0 \pi^+_{\rm s}$ decay, 
$Q \equiv M^\star - M - m_\pi$, 
obtained from the above technique is observed to have a narrow width, 
of order 190 k$e$V depending on $D^{0}$ decay mode, which is a combination of
the intrinsic width and detector resolution.  $M$ and $M^\star$ are the
reconstructed masses of the $D^0$ and $D^{\star +}$ candidates
respectively, and $m_\pi$ is the charged pion mass.  
In the mixing analyses described below, the distribution of
candidates in the $Q$--$M$ plane are fit to determine both RS 
and WS yields.

We calculate $t$ using only the vertical component of the $D^0$
candidate flight distance.  This is effective
because the vertical extent of the CESR luminous region has
$\sigma_{\rm vertical} = 7 \mu$m.  
The resolution on the $D^0$ decay point
($x_v$, $y_v$, $z_v$) is typically $40 \mu$m in each dimension.  
We express $t$ as 
$t = M/p_{\rm vertical} \times (y_v - y_b)/(c \tau_{D^0})$, where 
$p_{\rm vertical}$ is
the vertical component of the total momentum of the $D^0$ candidate.
The error in $t$, $\sigma_t$, is typically $0.4$ (in $D^0$ lifetimes), 
although when the $D^0$ direction is near the horizontal plane
$\sigma_t$ can be large.

\section{$CP$ Violation in $D^0$ Decay}\label{sec:kkpp}
$CP$ Violation in charm meson decay is expected to be small in the
Standard Model, which makes this a good place to look for
non-Standard Model effects.
Cabibbo-suppressed charm meson decays have all the necessary ingredients
for $CP$ violation -- multiple paths to the same final state and a
weak phase.  However, in order to get sizable $CP$ violation, the 
final state interactions need to contribute non-trivial phase shifts
between the amplitudes.  Large final state interactions may be
the reason why the observed ratio of branching ratios~\cite{PDG} 
$(D^0 \rightarrow K^+ K^-) / (D^0 \rightarrow \pi^+ \pi^-)$ 
is roughly twice the predicted value.  Thus, $D^{0}$ decays may
provide a good hunting ground for $CP$ violation.

Previous searches for mixing-induced~\cite{kpi} or direct~\cite{kspi0,PDG} 
$CP$ violation in the neutral charm meson system have set
limits of $\sim 30\%$ or a few percent, respectively.  We present results of
searches for direct $CP$ violation in neutral charm meson decay to pairs of
light pseudo-scalar mesons: $K^+ K^-$, $\pi^+ \pi^-$, $K^0_{\rm S} \pi^0$, $\pi^0
\pi^0$ and $K^0_{\rm S} K^0_{\rm S}$.

\subsection{Search for $CP$ Violation in $D^0 \rightarrow K^+ K^-$
and $D^0 \rightarrow \pi^+ \pi^-$ Decay}

The asymmetry result is obtained by fitting the energy release ($Q$) spectrum
of $D^{\star +} \rightarrow D^0 \pi^+_{\rm s}$ events for the yields
from $D^{0}$ and $\overline{D^{0}}$ decays.  The background-subtracted
$Q$ spectrum 
is fitted with a signal shape obtained from $K^+ \pi^-$ 
data and a background shape determined 
from Monte Carlo.  The $D^0$ mass
spectra are also fitted as a check.  

We measure the $CP$ asymmetry,
\[
A = \frac{\Gamma \left( D^0 \rightarrow f \right) - 
\Gamma \left( \overline{D^0} \rightarrow f \right)}
{\Gamma \left( D^0 \rightarrow f \right) +
\Gamma \left( \overline{D^0} \rightarrow f \right)},
\]
where the flavor of the $D^{0}$ is tagged by the slow pion charge in 
$D^{\star +} \rightarrow D^{0} \pi^+_{\rm s}$.
The decay asymmetry may be measured using this method because the
production and strong decay of $D^{\star \pm}$ are $CP$-conserving.

The parameters of the slow pion dominate the $Q$ distribution, 
so all modes have the same shape.
We do the fits in bins of $D^0$ momentum to eliminate bias
arising from
differences in the $D^0$ momentum spectra between the data and the MC.
The preliminary results are 
$A(K^+ K^-) = 0.0005 \pm 0.0218\ ({\rm stat}) \pm 0.0084\ 
({\rm syst})$ and $A(\pi^+ \pi^-) = 0.0195 \pm 0.0322\ ({\rm stat})
\pm 0.0084\ ({\rm syst})$.

We use many variations of the fit shapes, both empirical and 
analytical, to assess the systematic uncertainties due to the fitting 
procedure (0.69\%).
We also consider biases due to the detector material (0.07\%), the reconstruction
software (0.48\%), and forward--backward acceptance variations ($c \bar{c}$ 
pairs are not produced symmetrically in the forward/backward directions
in $e^+ e^-$ collisions at $\sqrt{s} \sim 10.6$ G$e$V, and the collision
point was not centered exactly in the middle of the detector) (0.014\%).

The measured asymmetries are consistent with zero, and no $CP$ violation
is seen.  These results are among the most precise
measurements~\cite{kspi0,OLDCP}.

\subsection{Search for $CP$ Violation in $D^0 \rightarrow K^0_{\rm
S} \pi^0$, $D^0 \rightarrow \pi^0 \pi^0$ and $D^0 \rightarrow K^0_{\rm S} 
K^0_{\rm S}$ Decay}

This analysis~\cite{jaffe} differs from the other analyses 
presented in this paper in 
some of its reconstruction techniques and in the data set used.  
The $\pi^0 \pi^0$ and
$K^0_{\rm S} \pi^0$ final states do not provide sufficiently precise
directional information about their parent $D^0$ to use the intersection of the $D^0$ 
projection and the CESR luminous region to refit the
slow pion, as described in Section~\ref{genmethod}. 
The $K^0_{\rm S} K^0_{\rm S}$ final state is treated the same for consistency.
This analysis uses the data from both the CLEO II and CLEO II.V
configurations
of the detector, corresponding to 13.7 ${\rm fb}^{-1}$ of integrated
luminosity. 

The $K^0_{\rm S}$ and $\pi^0$ candidates are reconstructed using only good 
quality tracks and showers.  The tracks (showers) whose combined 
invariant mass is close to the $K^0_{\rm S}$ ($\pi^0$) mass are
kinematically constrained to the $K^0_{\rm S}$ ($\pi^0$) mass, improving the 
$D^0$ mass resolution.  The tracks used to form $K^0_{\rm S}$
candidates are required to satisfy criteria designed to reduce
backgrounds from $D^0 \rightarrow \pi^+ \pi^- X$ decays and combinatorics.  
$D^0$ candidates with masses close to the known $D^0$ mass are selected to
determine the asymmetry $A$.  
Prominent peaks due to $D^{\star +} \rightarrow D^0
\pi^+_{\rm s}$ decay are observed in the $Q$ distributions of 
all three decay modes.

The yields of $D^0$ and $\overline{D^0}$ candidates for a given final
state are determined by subtracting the background yield obtained 
by fitting the sideband regions to a non-relativistic threshold function, 
$B(Q) = aQ^{1/2} + b Q^{3/2} + c Q^{5/2}$.  The background yield under
the signal is obtained by interpolation from the sidebands.

After background subtraction, we obtain $9099 \pm 153$ $K^0_{\rm S}
\pi^0$ candidates, $810 \pm 89$ $\pi^0 \pi^0$ candidates, and $65 \pm 14$ 
$K^0_{\rm S} K^0_{\rm S}$ candidates.  

We have searched for sources of false asymmetries introduced by the 
$\pi^+_{\rm s}$ finding (0.19\%),  fitting (0.5\%), 
and backgrounds (0.35\% $K^0_{\rm S} \pi^0$, 0\% $\pi^0 \pi^0$
and 12\% $K^0_{\rm S} K^0_{\rm S}$).
We find no significant bias, but apply the measured corrections and
add their uncertainties to the total.  We obtain the results 
$A(K^0_{\rm S} \pi^0) = (+0.1 \pm 1.3)\%$, 
$A(\pi^0 \pi^0) = (+0.1 \pm 4.8)\%$ and 
$A(K^0_{\rm S} K^0_{\rm S}) = (-23 \pm 19)\%$, 
where the uncertainties contain the 
combined statistical and systematic uncertainties.  All systematic
uncertainties, except for the 0.5\% uncertainty due to possible 
bias in the fitting method, are determined from data and would be
reduced in future higher luminosity samples.  

All measured asymmetries are consistent with zero, and no indication of
$CP$ violation is observed.  This measurement of 
$A(K^0_{\rm S} \pi^0)$ is a significant improvement over previous results, 
and the other two asymmetries reported are first measurements.

\section{First Rate Measurement of Wrong-Sign $D^0 \rightarrow K^+
  \pi^- \pi^0$ Decay} 

$D^{0} \rightarrow K\pi\pi^{0}$ candidates are reconstructed using the
selection criteria described in Section~\ref{genmethod}, with additional 
requirements specific to
this analysis.  In particular, $\pi^{0}$ candidates with momenta greater than
340~M$e$V/$c$ are reconstructed from pairs of photons detected in the
CsI crystal calorimeter.  Backgrounds are reduced by requiring
specific ionization of the pion and kaon candidates to be consistent
with their respective hypotheses.

We fit for the scale factor $S$ which relates the small number of
WS events, $N_{WS}$, to the large number $N_{RS}$ of RS events:
$N_{WS}=S\cdot N_{RS}$.  If the $U$-spin symmetry prediction of 
$SU(3)$-flavor is not badly violated~\cite{bib:GronauRosner}, we would expect
the WS channel to have a different resonant substructure than that observed
in the RS data.  We account for the efficiency difference which could
arise from this by allowing a correction factor $C$: $R = S\cdot C$.

\begin{figure}
\centerline{
 \psfig{file=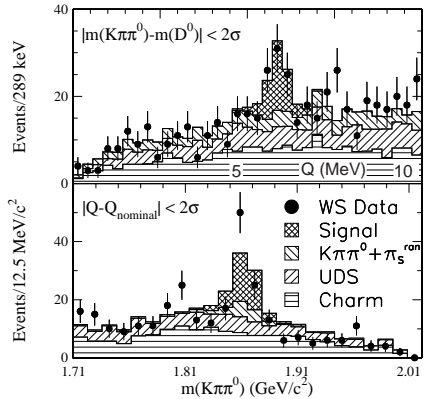,width=2.5in}
}
\caption{Results of the fit to the data to determine $S$.
  Projections in the variables $Q$ (top) and $M$ (bottom), after 
selecting the signal region (within two standard deviations) in the other
variable.}
\label{fig:alexmass}
\end{figure}

The scale factor $S$ is measured by performing a
two-dimensional maximum likelihood fit to the distribution in
$Q$ and $M$.  The signal distribution in these variables is
taken from the RS data.  The backgrounds are broken down into three
categories: 1) RS $\overline{D^{0}}\rightarrow K^{+}\pi^{-}\pi^{0}$ decay
combined with an uncorrelated $\pi_{s}$, 2) combinations from
$e^{+}e^{-}\rightarrow u\overline{u}$, $d\overline{d}$, and
$s\overline{s}$, and 3) combinations from charm particle
decays other than correctly reconstructed RS $\overline{D^{0}}$ decays.  The
background distributions are determined using a large Monte Carlo sample,
which corresponds to approximately eight times the integrated
luminosity of the data sample.  The $Q$--$M$ fit yields a
WS signal of  
$38 \pm 9$ events and a ratio $S=0.0043^{+11}_{-10}$.
Projections of the data and fit results in slices through the signal region in
each variable are shown in Fig.~\ref{fig:alexmass}.  
The statistical significance of this signal is found to be 4.9
standard deviations. 

The correction factor $C$ is determined
using a fit to the Dalitz plot of the WS data.  
The RS mode was recently fitted by CLEO~\cite{bib:bergfeld} and found
to have a rich Dalitz structure consisting  
of $\rho(770)^{+}$, $K^{\star }(892)^{-}$, $\overline{K^{\star }}(892)^{0}$, 
$\rho(1700)^{+}$, $\overline{K_{0}}(1430)^{0}$, $K_{0}(1430)^{-}$,
and $K^{\star }(1680)^{-}$ resonances and non-resonant contributions.
In this fit, the amplitudes and
phases are initialized to the RS values, with those
corresponding to the $K^{\star }(892)^{+}$ and $K^{\star }(892)^{0}$ resonances 
allowed to vary relative to the dominant $\rho(770)^{-}$ and other minor
contributions.  
We measure a correction of $C=1.00\pm 0.02\ ({\rm stat})$.  
Studies are under way to examine the extent and significance of the
surprising similarity between the RS and WS Dalitz plots.

We estimate an uncertainty of 14\% on $S$ due to the Monte Carlo
$Q$--$M$ background, based on a series of fits using
specific background subregions of the $Q$--$M$ plane.  
The 9.5\% uncertainty in $C$ has contributions from the 
unknown amplitudes and phases that are fixed in the determination of
$C$ (8\%), the Dalitz plot fit method (3.6\%), and the Dalitz plot of 
non-$\overline{D^{0}}$ backgrounds (3\%).
Other minor contributions come from
mismodeling of selection variables (3\%), and statistics of the Monte
Carlo background sample (2.4\%). 

We measure the WS rate to be $R = 0.0043^{+0.0011}_{-0.0010}\
({\rm stat}) \pm 0.0007\ ({\rm syst})$ (preliminary). 
Work is in progress to use the lifetime distribution
of this sample to yield independent limits on $x^{\prime}$,
$y^{\prime}$, and $R_{DCSD}$.
%

%
\section{Search for $CP$-Dependent Lifetime Differences Due to 
$D^0-\overline{D^0}$ Mixing}

In the limit of no $CP$ violation in the neutral $D$ system,
we can express $y$ as
\begin{equation}
y = \frac{\tau_{\overline{CP}}}{\tau_{CP+}} - 1 ,
\end{equation}
where $\tau_{\overline{CP}}$ is the lifetime of a $CP$ neutral state, such
as $K^{-}\pi^{+}$, and $\tau_{CP+}$ is the lifetime of a $CP$ even state,
such as $K^{+}K^{-}$ or $\pi^{+}\pi^{-}$.  Thus, to measure $y$ we simply
measure the ratio of the lifetimes of $D^{0} \to K^{-}\pi^{+}$ and $D^{0}
\to K^{+}K^{-}$ or $\pi^{+}\pi^{-}$.  Since the final states are very
similar, our backgrounds  
are small, and cross-feed among the final states is negligible, many
of the sources of uncertainty cancel in the ratio.  A similar analysis
has recently been published by FOCUS comparing the
$K^{-}\pi^{+}$ and $K^{+}K^{-}$ final states~\cite{FOCUSKK}.

We fit the proper time distributions of candidates near the $D$ mass
using an unbinned maximum likelihood fit.  The proper time measurement
technique is described in Section~\ref{genmethod}.

The signal likelihood function is 
an exponential convolved with three resolution Gaussians.  The 
width of the primary Gaussian is due to the propagation of
errors from the track fit to the flight distance and momentum
for each candidate.  The second and third Gaussians represent candidates
that have been mismeasured by the addition of spurious tracking hits or
due to hard non-Gaussian scatters in the material of the detector.
The first of these has its width determined in the fit to the copious 
$K^{-}\pi^{+}$ sample, while the second is fixed to a large value of 8~ps.
The relative contribution of the two mismeasured signal resolutions
is determined using the $K^{-}\pi^{+}$ data sample and fixed for the 
$K^{+}K^{-}$ and $\pi^{+}\pi^{-}$ samples.
According to our simulation, the fraction of the second Gaussian
is about 4\% of the well-measured signal and the third Gaussian
is less than 0.1\% of the signal.  The probability for a candidate to be signal
is determined by its measured mass $M$, and is based on a fit to  
that distribution.

The background is considered to have contributions with both zero 
and non-zero lifetimes.
All parameters that describe the background are allowed to vary in the fits 
except for the width of the widest Gaussian which is fixed to 8 ps.

The background in all three samples has a large component
with a lifetime consistent with that of the $D^0$.  This agrees
with the prediction of our simulation that the background with 
lifetime is dominated by misreconstructed fragments of charm decays.

We calculate $y$ separately for the $K^{+}K^{-}$ and $\pi^{+}\pi^{-}$
samples.  Systematic uncertainties are dominated by the statistical uncertainty
in a Monte Carlo study used to determine small corrections, consistent with
zero, that are applied to the measured result to account for differences 
between measured and generated values of the lifetimes ($\pm 0.009$).  
Additional significant systematic uncertainties
come from variations in the description of the background ($\pm 0.008$),
uncertainties in our model of the proper time resolution ($\pm 0.005$), and
details of the fit procedure ($\pm 0.005$), where the listed values are the
contribution to the average result.
Our preliminary results are $y_{KK} = -0.019 \pm 0.029\ ({\rm stat})
\pm 0.016\ ({\rm syst})$ and $y_{\pi\pi} = 0.005 \pm 0.043\ ({\rm stat})
\pm 0.018\ ({\rm syst})$. 
We form a weighted average of the two to get
$y =  -0.011 \pm 0.025\ ({\rm stat}) \pm 0.014\ ({\rm syst})$ (preliminary),
which is consistent with zero.  It is also consistent 
with our previous result using
$D^{0}\rightarrow K^{+}\pi^{-}$~\cite{kpi} and the FOCUS results using 
$D^{0}\rightarrow K^{+}\pi^{-}$~\cite{FOCUS} and $D^{0}\rightarrow
K^{+}K^{-}$~\cite{FOCUSKK}.  

\section{\dstarp\ Width Measurement}

The challenge of measuring the width of the $D^{\star +}$ is
understanding the experimental resolution, which exceeds the width
we are trying to measure. 
Unfortunately, there is no decay mode with small width, large
cross-section, and similar kinematics to use as a calibration of the
method.  We depend on exhaustive comparisons between our
detector simulation and data in order to
understand the effect 
of candidates with mismeasured hits, errors in pattern recognition, or
large angle Coulomb scattering on the measured width.  

In addition to the cuts described in Section~\ref{genmethod}, we require
events to be within the kinematically allowed 
regions of \dstarp\ momentum, $\pi_{s}^{+}$ momentum, and
$D^{0}$-$\pi_{s}^{+}$ opening angle in order to remove a small
amount of misreconstructed background.  We also remove poorly measured
events by requiring $\sigma_Q < 200$ k$e$V and do not consider
 $D^0$ candidates within 0.3 radians of the 
horizontal plane.

We assume that the intrinsic width of the $D^0$ is negligible,
$\Gamma(D^0) \ll \Gamma(D^{\star +})$, implying that the width of $Q$
is simply a convolution of the shape given by the $D^{\star +}$ width and
the tracking system response function.  We
consider pairs of $Q$ and $\sigma_Q$ for the decay
$D^{\star +} \to \pi_{s}^+ D^0 \to K^-\pi^+\pi_{s}^+$,
where $\sigma_Q$ is given for each candidate by propagating the
tracking errors from the kinematic fit of the charged tracks.

The width is extracted using an unbinned maximum likelihood fit to the
$Q$ distribution.  The observed $Q$ distribution is
fitted to a Breit-Wigner signal and polynomial background, with its shape
fixed by the simulation.

\begin{figure}
  \centerline{
  \epsfig{file=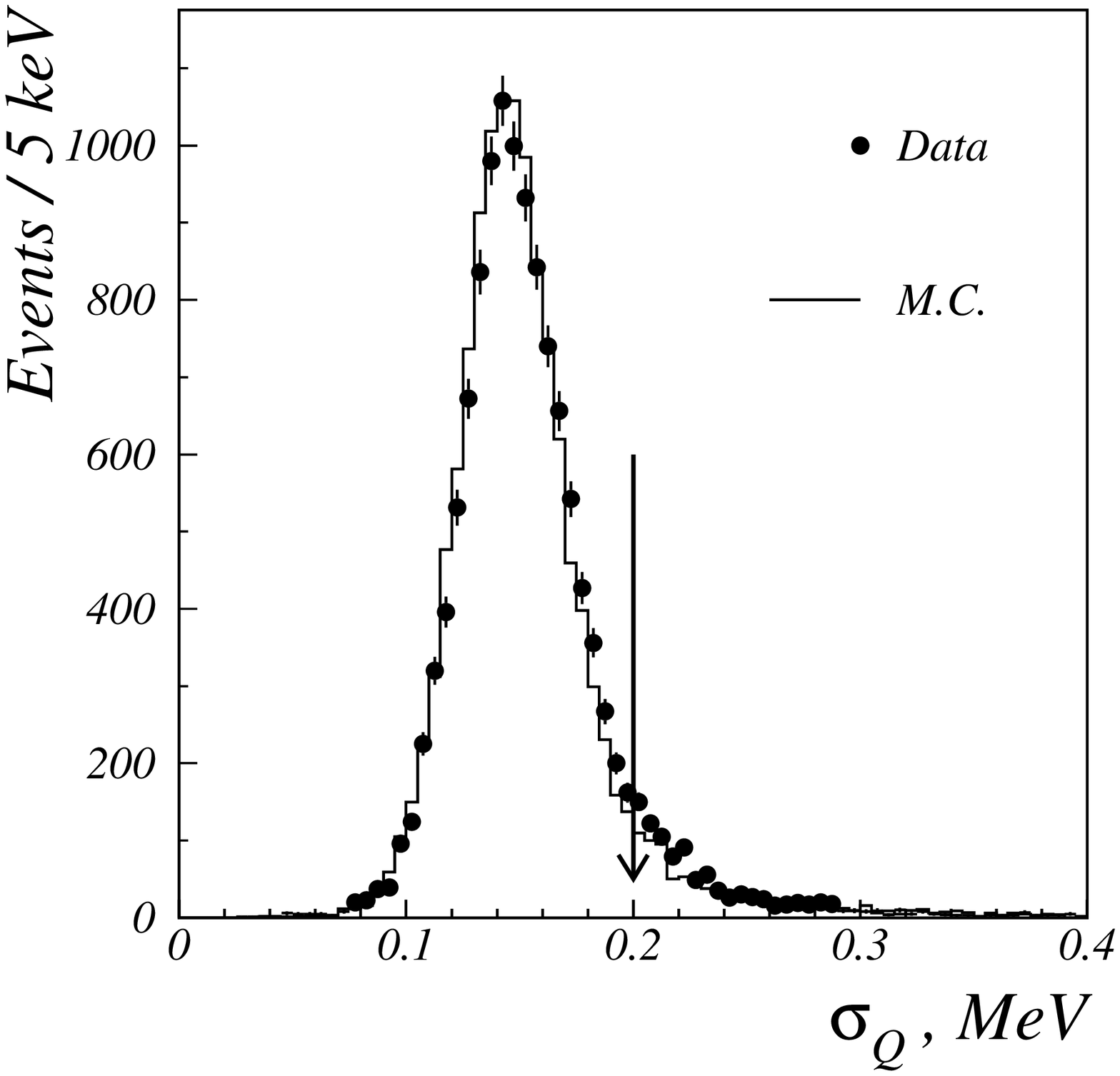,width=1.75in}
  \epsfxsize=1.75in
  \epsfbox{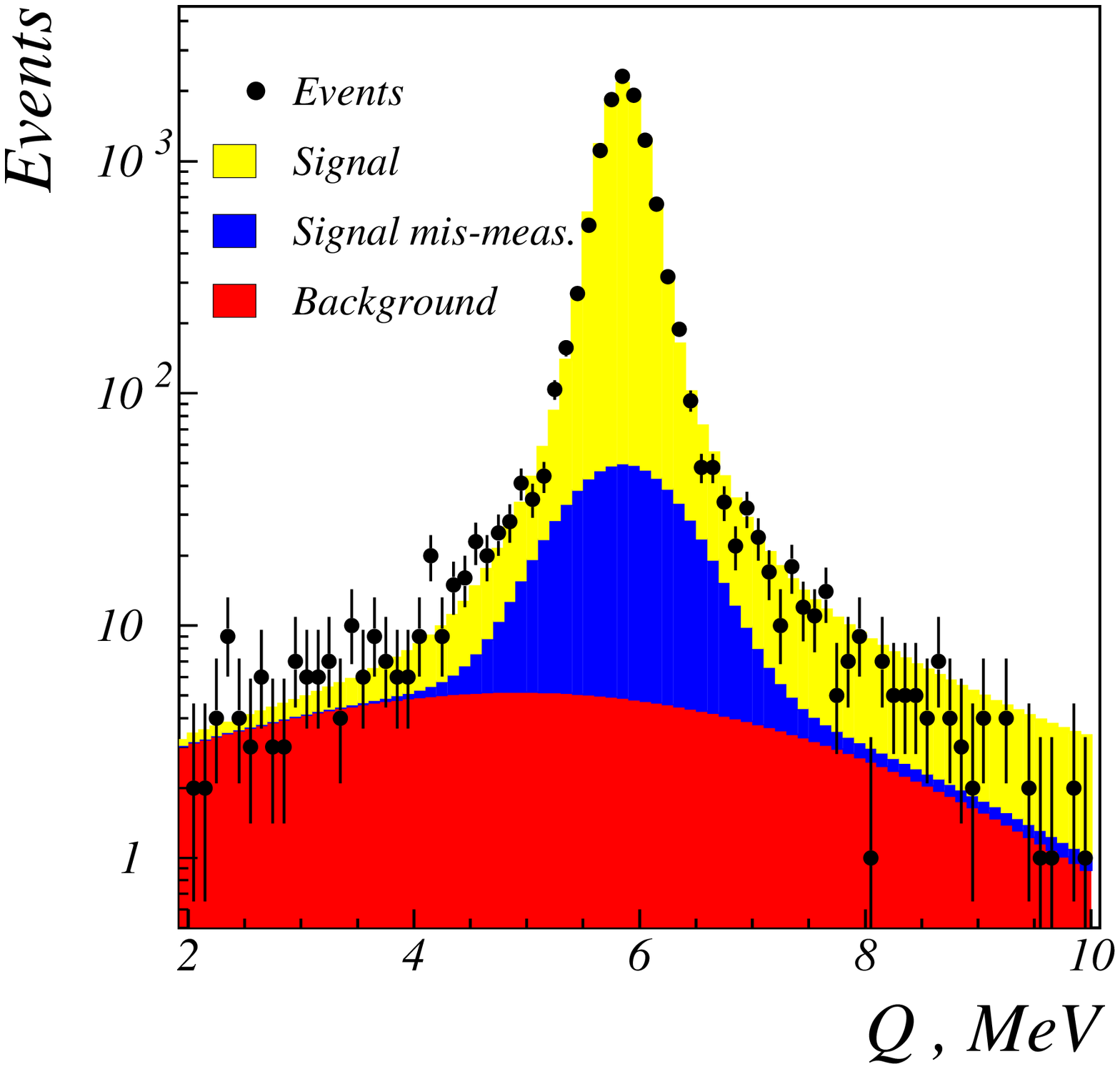}
  }
  \caption{\label{fig:errorcompare} Distribution of $\sigma_Q$,
    the uncertainty on $Q$ as determined from propagating
    track fitting errors (left).  Fit to the data sample (right).}
\end{figure}

        For each candidate, the signal shape is convolved with
a  Gaussian resolution function with width given by the event-by-event
error on $Q$, $\sigma_Q$.  The detector resolution on $Q$ is
approximately 150~k$e$V.
We find excellent agreement between the
simulation of $\sigma_Q$ and that observed in the data, as shown in
Fig.~\ref{fig:errorcompare}. This reflects the correct modeling of the 
kinematics and sources of errors on the tracks in the simulation, such
as number of hits on a track and the effects of multiple scattering in
detector material.  

In order to account for poorly measured events, we allow a small fraction
$f_{mis}$ of the signal  to be parameterized 
by a single Gaussian resolution function of width $\sigma_{mis}$.
In the fit, we constrain the level of this contribution to $f_{mis}=5.3 \pm
0.5$\%.  This value was determined by fitting a range of simulated
$D^{\star +}$ widths, while allowing $\sigma_{mis}$ to vary. 

The fitter has been tested extensively using Monte Carlo samples
generated with different $\Gamma(D^{\star +})$ and is found to
reproduce inputs ranging from 0 to 130~k$e$V.  In this study, we observe a bias
of $-2.7\pm2.1$~k$e$V, which is consistent with zero.  We apply this
small correction to our fit result.

We observe a corrected width of $\Gamma(D^{\star +}) = 96.2 \pm 4.0\
({\rm stat})\ {\rm k}e{\rm V}$.


We test the simulation of the $Q$ distribution
and estimate the corresponding systematic uncertainty by comparing its
dependence on kinematic variables of the \dstarp\ decay with data.
Specifically, we compare the Gaussian peak width and mean as a function of
$P_{D^{0}}$, $P_{\pi_{s}^{+}}$, $\theta_{D^{0}-\pi_{s}^{+}}$, 
$\partial Q/\partial P_{D^{0}}$, $\partial Q/\partial P_{\pi_{s}^{+}}$, and
$\partial Q/\partial\theta_{D^{0}-\pi_{s}^{+}}$.  The derivatives test
correlations among the basic kinematic variables.
We compare by dividing the sample into ten slices in each variable and
fitting the ten distributions of $Q$ to Gaussians.

We observe some discrepancy between the Gaussian mean of the $Q$ peak
in data and Monte Carlo as a function of $P_{\pi_{s}^{+}}$, 
$\partial Q/\partial P_{\pi_{s}^{+}}$, and $\partial Q/\partial 
P_{D^{0}}$.  While measuring the mean $Q$ is not our goal, we include
the observed deviation as a systematic uncertainty of 16 k$e$V on  
$\Gamma(D^{\star +})$.

We find excellent agreement between the Gaussian width of the $Q$
peak in data and Monte Carlo as a function of all variables if events
are generated with an intrinsic width in the range $\Gamma(D^{\star +})
= 90-100$~k$e$V.  Even when comparing with a sample of zero intrinsic
width, we see excellent agreement in the dependence on the kinematic
variables.

In order to test our modeling of effects that contribute to the
tracking errors, we perform fits in which we allow a scale factor $k$ to 
multiply the event-by-event error $\sigma_{Q}$.
We estimate this effect by varying our cut on $\sigma_{Q}$ from the 
nominal 200~k$e$V in the range 75~k$e$V to 400~k$e$V.  Several
effects, such as improper modeling of material in 
the detector, could lead to deviations of $k$ from unity.  Repeating 
our analysis with all parameters fixed except $k$, we find $k=1.00
\pm 0.04$.  We measure an uncertainty of $\pm 11$~k$e$V on 
$\Gamma(D^{\star +})$ by rerunning the analysis with $k$ fixed at its
limits.

We take into account correlations among the less well-measured
parameters of the fit, such as $k$, $f_{mis}$, and $\sigma_{mis}$,
by varying each parameter one standard deviation from its central
fit value.  We find an uncertainty of $\pm 8$~k$e$V.

Biases in the reconstruction are estimated by replacing
reconstructed parameters with their generated values in the analysis.
We find only a small bias in the reconstruction of the $D^0$ origin
point.  This contributes an uncertainty $\pm 4$ k$e$V.

Uncertainties from the background shape are determined by allowing the
coefficients of the background polynomial to vary in the fit.
We observe a change on the width of $\pm 4$ k$e$V.

These studies were confirmed using subsamples of the data. 
One sample uses events restricted to kinematic
regions in which the mean $Q$ and $\sigma_{Q}$ are well-modeled:
$|\partial Q/\partial P_{D^{0}}| \leq 0.005$ and 
$|\partial Q/\partial P_{\pi_{s}^{+}}| \leq 0.05$.  The other
uses events in which very tight selection criteria are applied to the
tracks.  In both cases the background from poorly measured tracks is
negligible and not included in the fit.  We measure 
$\Gamma(D^{\star +})$ to be consistent with our nominal sample:
($103.8\pm 5.9\ ({\rm stat})$)~k$e$V 
(($104\pm 20\ ({\rm stat})$)~k$e$V) in the former (latter) sample.

        We measure the width of the $D^{\star +}$ to be
$\Gamma(D^{\star +}) = (96 \pm 4\ ({\rm stat}) \pm 22\ ({\rm syst}))\
{\rm k}e{\rm V}$ (preliminary)
by studying the
distribution of the energy release in $D^{\star +} \to D^0 \pi^+$ followed
by $D^0 \to K^- \pi^+$ decay.  

This is the first measurement of the $D^{\star +}$ width, corresponding
to a strong coupling~\cite{pred} of
$g = 17.9 \pm 0.3\ ({\rm stat}) \pm 1.9\ ({\rm syst})$ (preliminary).
This is consistent with theoretical predictions based on HQET and
relativistic quark models, but higher than predictions based on QCD
sum rules.  

\section{Summary}
We  present preliminary results of several analyses 
searching for the effects of $CP$ violation and mixing 
in the decay of $D^0$ mesons.  
We find no evidence of $CP$ asymmetry in five different 
two-body decay modes of the $D^0$ to pairs of light pseudo-scalar mesons:
$A_{CP}(K^+ K^-) = (0.05 \pm 2.18 \pm 0.84)\%$, 
$A_{CP}(\pi^+ \pi^-) = (2.0 \pm 3.2 \pm 0.8)\%$,
$A_{CP}(K^0_{\rm S} \pi^0) = (+0.1 \pm 1.3)\%$, 
$A_{CP}(\pi^0 \pi^0) = (+0.1 \pm 4.8)\%$ and 
$A_{CP}(K^0_{\rm S} K^0_{\rm S}) = (-23 \pm 19)\%$.
We present the first measurement of the rate of WS 
$D^0 \rightarrow K^+ \pi^- \pi^0$ decay:
$R = 0.0043^{+0.0011}_{-0.0010} \pm 0.0007$.
We describe a measurement of the mixing 
parameter $y={\Delta\Gamma\over 2 \Gamma}$ 
by searching for a lifetime difference between the $CP$ 
neutral $K^+ \pi^-$ and the 
$CP$ even $K^+K^-$ and $\pi^+\pi^-$ final states.  
Under the assumption that $CP$ is conserved we find 
$y = -0.011 \pm 0.025 \pm 0.014$.
Finally, we describe our preliminary measurement of the \dstarp\ intrinsic 
width of $\Gamma(D^{\star +}) = (96 \pm 4\ ({\rm stat}) \pm 22\ ({\rm syst}))\
{\rm k}e{\rm V}$.

\section{Acknowledgments}
We gratefully acknowledge the effort of the CESR staff in providing us with
excellent luminosity and running conditions.
M. Selen thanks the PFF program of the NSF and the Research Corporation, 
and A.H. Mahmood thanks the Texas Advanced Research Program.
This work was supported by the National Science Foundation, the
U.S. Department of Energy, and the Natural Sciences and Engineering Research 
Council of Canada.


\begin{thebibliography}{99}
\bibitem{charge} Charge conjugation is implied throughout, except
  in discussions of $CP$ violation.
\end{thebibliography}
\end{document}